\documentclass[pra,showpacs,superscriptaddress,amssymb,twocolumn,bibliography,nofootinbib]{revtex4-2} 

\usepackage[utf8]{inputenc}
\usepackage{graphicx}
\usepackage{amssymb,amsmath,amsthm,amsfonts}
\usepackage{hyperref}
\usepackage{bm}   
\usepackage{subfloat}
\usepackage{float}
\usepackage[caption=false,justification=justified]{subfig}
\usepackage{xcolor}

\newcommand{\revision}[1]{\textcolor{black}{#1}}

\usepackage[normalem]{ulem}

\usepackage[utf8]{inputenc}

\usepackage{float}
\usepackage{amsmath,amssymb,setspace}
\usepackage{graphicx}
\usepackage{bm, bbm}      
\usepackage{srcltx}
\usepackage{upgreek}
\usepackage{slashed}
\usepackage{dirtytalk}
\usepackage{mathtools}
\usepackage{titlesec}
\usepackage{titlecaps}
\usepackage{bm, bbm} 
\usepackage{srcltx}
\usepackage{slashed}

\newcommand{\Gate}[1]{\textsc{#1}}
\newcommand{\hgate}{\Gate{h}}

\newcommand{\sgate}{\Gate{s}}

\newcommand{\rzgate}{{\Gate{r}}_{z}}
\newcommand{\rygate}{{\Gate{r}}_{y}}
\newcommand{\rxgate}{{\Gate{r}}_{x}}

\newcommand\bea{\begin{eqnarray}}
\newcommand\eea{\end{eqnarray}}

\newcommand{\Bra}[1]{\langle#1|} 
\newcommand{\Ket}[1]{|#1\rangle}

\def\=d{\, {\buildrel \rm def  \over =} \,}

\def\sqr#1#2{{\vcenter{\vbox{\hrule height.#2pt \hbox{\vrule width.#2pt height#1pt \kern#1pt \vrule width.#2pt}\hrule height.#2pt}}}}

\def\beq#1{\begin{equation} \label{#1}}
\def\eeq{\end{equation}}
\def\ben{\begin{equation*}}
\def\een{\end{equation*}}
\def\bequa{\begin{eqnarray}}
\def\eequa{\end{eqnarray}}

\def\bf#1{\bm{#1}}

\def\=d{\, {\buildrel \rm def  \over =} \,}

\def\sqr#1#2{{\vcenter{\vbox{\hrule height.#2pt \hbox{\vrule width.#2pt height#1pt \kern#1pt \vrule width.#2pt}\hrule height.#2pt}}}}

\def\beq#1{\begin{equation} \label{#1}}
\def\eeq{\end{equation}}
\def\ben{\begin{equation*}}
\def\een{\end{equation*}}
\def\bequa{\begin{eqnarray}}
\def\eequa{\end{eqnarray}}

\def\bf#1{\bm{#1}}

\def\beq{\begin{equation}}
\def\eeq{\end{equation}}
\def\d{\partial}

\def\e{{\rm e}}

\def\d{\partial}
\def\l{\left(}
\def\r{\right)}

\usepackage{siunitx}

\begin{document}

\title{\revision{Optimizing Electronic Structure Simulations on a Trapped-ion Quantum Computer using Problem Decomposition}}

\author{Yukio Kawashima}
\email{snowbirth@gmail.com}
\affiliation{1QB Information Technologies (1QBit) \\200-1285 Pender St W, Vancouver, BC, V6E 4B1, Canada}

\author{Erika Lloyd}
\email{erika.lloyd@1qbit.com}
\affiliation{1QB Information Technologies (1QBit) \\200-1285 Pender St W, Vancouver, BC, V6E 4B1, Canada} 
 
\author{Marc P. Coons}
\email{MPCoons@dow.com}
\affiliation{Dow, Core R\&D, Chemical Science, 1776 Building, Midland, MI, 48674, USA}

\author{Yunseong Nam}
\email{nam@ionq.co}
\affiliation{IonQ, 4505 Campus Drive, College Park, MD, 20740, USA}

\author{\\ Shunji Matsuura}
\affiliation{1QB Information Technologies (1QBit) \\200-1285 Pender St W, Vancouver, BC, V6E 4B1, Canada}

\author{Alejandro J. Garza}
\affiliation{Dow, Core R\&D, Chemical Science, 1776 Building, Midland, MI, 48674, USA}

\author{Sonika Johri}
\affiliation{IonQ, 4505 Campus Drive, College Park, MD, 20740, USA}

\author{Lee Huntington}
\affiliation{1QB Information Technologies (1QBit) \\200-1285 Pender St W, Vancouver, BC, V6E 4B1, Canada}

\author{Valentin Senicourt}
\affiliation{1QB Information Technologies (1QBit) \\200-1285 Pender St W, Vancouver, BC, V6E 4B1, Canada}

\author{\\Andrii O. Maksymov}
\affiliation{IonQ, 4505 Campus Drive, College Park, MD, 20740, USA}

\author{Jason H. V. Nguyen}
\affiliation{IonQ, 4505 Campus Drive, College Park, MD, 20740, USA}

\author{Jungsang Kim}
\affiliation{IonQ, 4505 Campus Drive, College Park, MD, 20740, USA}

\author{Nima Alidoust}
\affiliation{1QB Information Technologies (1QBit) \\200-1285 Pender St W, Vancouver, BC, V6E 4B1, Canada}

\author{Arman Zaribafiyan}
\affiliation{1QB Information Technologies (1QBit) \\200-1285 Pender St W, Vancouver, BC, V6E 4B1, Canada}

\author{Takeshi Yamazaki}
\email{takeshi.yamazaki@1qbit.com}
\affiliation{1QB Information Technologies (1QBit) \\200-1285 Pender St W, Vancouver, BC, V6E 4B1, Canada}

\begin{abstract}
    \begin{center}
        (Date: \today)
        \vspace{-.9em}
        \section*{Abstract}
        \vspace{-.9em}
    \end{center}
\revision{Quantum computers have the potential to advance material design and drug discovery by  performing costly electronic structure calculations. A critical aspect of this application requires optimizing the limited resources of the quantum hardware.}  Here, we experimentally demonstrate an end-to-end pipeline that focuses on minimizing quantum resources while maintaining accuracy. Using density matrix embedding theory as a problem decomposition technique, and an ion-trap quantum computer, we simulate a ring of 10 hydrogen atoms without freezing any electrons. The originally 20-qubit system is decomposed into 10 two-qubit problems, making it amenable to currently available hardware. Combining this decomposition with a qubit coupled cluster circuit ansatz, circuit optimization, and density matrix purification, we accurately reproduce the potential energy curve in agreement with the full configuration interaction energy in the minimal basis set. Our experimental results are an early demonstration of the potential for problem decomposition to accurately simulate large molecules on quantum hardware.

\end{abstract}

\maketitle

\titleformat{\section}
  {\large\bfseries\center}{\thesection.}{10pt}{\MakeUppercase}{}

\revision{\section*{Introduction}}

Electronic structure simulation is an essential tool
for understanding chemical properties of molecules. 
It is a basis for contemporary materials design and drug discovery.
Performing accurate electronic structure simulations on classical computers requires a great amount of computational resources. In particular, they grow exponentially with the system size when employing the full configuration interaction (full CI) method, which calculates the  exact solution of the electronic Schr\"{o}dinger equation in a given basis set.  Quantum computing, a computing paradigm that leverages the
laws of quantum physics, \revision{has the potential} to deliver scalable and accurate
electronic structure calculations~\cite{Manin:1980, Feynman:1982}
beyond the reach of classical computers.

Quantum computing technologies are rapidly advancing,
and there has been major progress in simulating molecular systems in the last two \mbox{decades~\cite{Aspuru-Guzik:2005, Peruzzo:2014, OMalley:2016, Shen:2017, Kandala:2017, Hempel:2018, 
Nam:2019, McCaskey:2019, Rice:2020, Stober:2020, Arute:2020}}. However, simulating the electronic structure of industrially relevant molecular systems on today's noisy, intermediate-scale quantum (NISQ) devices~\cite{Preskill:2018} will require systematically scalable, robust methods that allow a given problem to be represented by a small number of qubits and shallow quantum circuits. The treatment of electron correlation is also necessary for applying electronic structure calculations to make accurate predictions about the process of a chemical reaction. Calculating the molecular energy remains a challenge, even for small systems, without limiting the number of configurations or the number of electrons active in performing the electronic structure calculations. \revision{Without being able to treat correlations in a scalable way,} the stringent constraints of NISQ devices will inhibit their ability to perform high-accuracy simulations of larger molecular systems, particularly those systems where the electron correlation is strong.

\revision{Progress is being made on these issues as hardware develops.} The largest calculation of the total energy including electron correlation to date is the simulation of BeH$_{2}$ using six qubits for the six-electron problem~\cite{Kandala:2017}. Note further that some of the authors of the present manuscript \revision{simulated} a water molecule on a trapped-ion quantum computer. Using a small number of electron configurations that are known to contribute significantly to the total energy of the molecule, the energy estimates obtained were within the widely used measure for chemical accuracy ($1.5936 \times 10^{-3}$ hartrees) when compared to classically simulated results~\cite{Nam:2019}. Other recent research~\cite{Arute:2020} simulated a chain of 12 hydrogen atoms using 12 qubits and the Hartree--Fock method. This is a system with 12 uncorrelated electrons and set the record in terms of the largest number of qubits used for a chemistry simulation.

\revision{To approach the simulation of larger molecules,} problem decomposition techniques can be used to decompose a given molecular system into small subsystems, without sacrificing the accuracy of the electronic structure calculation for a wide class of chemical systems~\cite{Collins:2015, Raghavachari:2015, Sun:2016, Neugebauer:2018}. These techniques admit a more compact representation of a molecule, \revision{enabling} the explicit inclusion of more electrons in calculating correlation energies. Although the amount of reduction in the computational cost and resulting accuracy is dependent on the problem decomposition algorithm and the system being studied, these techniques have the potential to substantially reduce the qubit count requirements in electronic structure simulations~\cite{Rubin:2016, Bauer:2016, Reiher:2017, Yamazaki:2018, Kuhn:2018, Gao:2019, mochizuki:2019, Verma:2020, Takeshita:2020}. 

  For our experiment, we look at a ring of 10 hydrogen atoms taking all electrons into account and without using the frozen core approximation. We choose density matrix embedding theory ~\cite{Knizia:2012,Wouters:2016}  as the appropriate problem decomposition method based on previous studies using classical simulations for this system~\cite{Wouters:2016}, and for the method's success in quantum simulations of the Hubbard model~\cite{Rubin:2016}. DMET has been studied for molecular systems, ranging from model systems such as a ring or a lattice of hydrogen atoms~\cite{Knizia:2013aa} to more-realistic organic molecules~\cite{Wouters:2016,Pham:2018aa}.  Despite the strength and wide applicability of DMET-based decomposition, more complex molecular systems might require a different choice of problem decomposition technique.
  
In this work we demonstrate experimentally how, given a limited fragment size, we can extend the applicability of quantum hardware to larger systems with the aid of classical computations \revision{that implement problem decomposition. We do this by applying the DMET method to generate fragments, and solve the electronic structure problem of the fragments using a quantum algorithm}. The most appropriate algorithm for current hardware is the variational quantum eigensolver (VQE) ~\cite{Peruzzo:2014} whose advantage is enabling shallow circuits that can realistically be executed, at the expense of taking more measurements. We use the established qubit coupled cluster method ~\cite{Ryabinkin:2018} to generate the parametric ansatz required for running the VQE algorithm. To run our circuits we use a trapped-ion quantum computer~\cite{Shen:2017,Hempel:2018,Nam:2019},
a platform that allows for an efficient implementation of quantum
circuits via complete qubit connectivity~\cite{Linke_2017,Nam_2019_FC,Grzesiak_2020}. We further employ a density matrix purification algorithm to post-process the experimentally determined results and mitigate residual error. \revision{Our results demonstrate the success of an end-to-end pipeline using problem decomposition to accurately solve a molecular problem.}

\begin{figure*}[hbtp]
\centering
\includegraphics[width=0.9\textwidth]{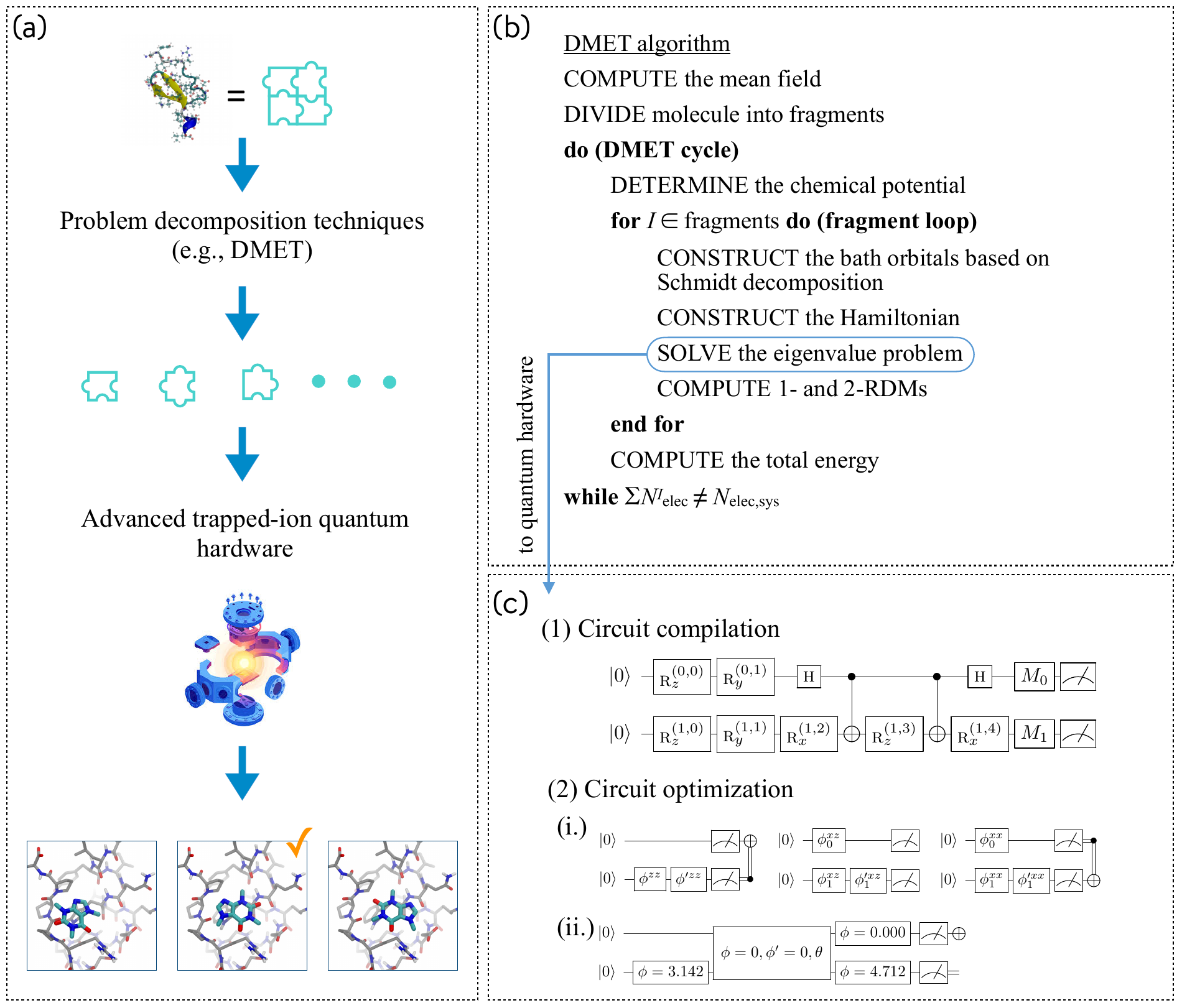}
\caption{Problem decomposition--based pipeline for efficient electronic structure simulation on a quantum computer. \mbox{(a) Schematic} illustration of the pipeline. (b) The density matrix embedding theory (DMET) algorithm. ``1- and 2-RDMs'' refers to one- and two-particle reduced density matrices. (c) Pre- and post-optimizing compilation circuits. (1) Pre-optimizing compilation input, with the gates $M_0$ and $M_1$ chosen appropriately for the different measurement bases. (2) Post-optimizing compilation output. (i.) The three output circuits for ZZ (left), XZ (middle), and XX (right).  Note that XZ and ZX result in the same circuit, outside of the relabelling of the qubit indices. See Tables~\ref{tab:param-pre} and \ref{tab:param-post} in Methods section for the numerical values of the gate parameters. (ii.) Post-optimization circuits for YY, required for the classical post-processing of our simulation data. See Table~\ref{tab:param-post2} in Methods section for the numerical values of the gate parameters.}
\label{fig:pipeline}
\end{figure*}

\section*{Results and Discussion}
\subsection*{\revision{Numerical results of the density matrix embedding theory and the qubit coupled cluster methods}}
We begin by describing our DMET-based methodology for simulating the electronic structure of a molecular system (see Methods section for further details).
Consider a system described by a second-quantized Hamiltonian
$\hat{H} = H^{1\text{-e}} + H^{2\text{-e}}$,
where $H^{1\text{-e}}$ is the one-electron interaction
and $H^{2\text{-e}}$ is the two-electron interaction of
the entire molecule.
In contrast to simulating the entire system using $\hat{H}$, in DMET, the system to be simulated is divided into small fragments. Each fragment is treated as an open quantum system, entangled with its surrounding environment, or bath. 
Here we use the mean-field, Hartree--Fock (HF) solution of the entire molecular system as a pre-processing step to find local orbitals that we use to fragment the molecule. Then, the following iterative process, which we call the DMET cycle, initiates. Once the cycle terminates, the electronic structure calculation of the entire molecule is complete. The DMET algorithm is shown in Fig.~\ref{fig:pipeline}(b).

The DMET cycle begins by constructing the bath orbitals for each fragment. The bath orbitals describe the environment that is active for the electronic structure calculation of the fragment, by virtue of Schmidt decomposition~\cite{Peschel:2012}. Note that, if the bath is large, the description of the bath can be greatly simplified. With the simple description of the bath, the Hamiltonian $H^\text{{A}}$ for each fragment A, along with its specific bath B, is constructed according to the equation
\begin{equation}
\label{eq:embed_hamiltonian}
\begin{split}
H^{\text{A}}= H^{1\text{-e,AB}} + H^{2\text{-e,A}} - \mu N^\text{A},
\end{split}
\end{equation}
where $H^{1\text{-e,AB}}$ denotes the one-electron interaction within
and across the fragment and the bath, $H^{2\text{-e,A}}$ denotes the two-electron
interaction within the fragment, $\mu$ is the chemical potential, and 
$N^\text{A}$ is the number of electrons in the fragment. See Methods section for more details.
We use a quantum computer in the DMET calculation to accurately evaluate the minimal expectation value of $H^\text{{A}}$, as well as the number of electrons $N^\text{A}$ in fragment A. 

Once both the energy expectation value and the number of electrons for each fragment are computed, we combine them to compute the total system energy, and check for self-consistency. In particular, we choose to compute the sum of the number of electrons in each fragment and check whether the sum is equal to the total number of electrons in the system. If the sum is within a pre-specified range with respect to the total number of electrons, the DMET cycle terminates. If the sum is not within the specified range, we run the DMET cycle again, with the chemical potential $\mu$ updated as the difference between the sum and the total number of electrons. 

Our explicit example of calculating the electronic structure of a molecule using DMET, 
 \revision{is to} simulate a ring of 10 hydrogen atoms, H$_{10}$. \revision{For this molecule we take advantage of its symmetry to create identical subproblems. This allows us to use a single fragment to solve for the entire system. If this symmetry was not present, we would need to solve each subproblem individually which could be done in parallel. To minimize quantum resources, we choose to divide the molecule into 10 one atom fragments, allocating two spin-orbitals for the fragment and for the bath.} This may be compared to a total of 20 spin-orbitals in the simulation of the entire molecule,
which shows a large reduction in the problem size: a 20-qubit problem is reduced to 10 two-qubit problems.

\revision{Classical simulations show that this decomposition reproduces full CI energy within chemical accuracy in all regions of the dissociation curve, except for the repulsive wall. This is consistent with the studies in~\cite{Wouters:2016}, which also show how correlations can be added to improve the approximation by increasing the fragment size, or by adding an additional self-consistency loop to optimize the correlation potential in the DMET approach.  For the points along the dissociation curve that we will explore experimentally, we calculate the total energy per atom of the DMET fragments with a full CI solver, and denote the results under DMET-FCI in Table~\ref{fig:dmet_energies}. }

To estimate the expectation value of the fragment energy and the number of particles per fragment \revision{on a quantum device}, we use VQE~\cite{Peruzzo:2014} with the qubit coupled-cluster (QCC) ansatz~\cite{Ryabinkin:2018}. For all calculations we use the symmetry-conserving Bravyi--Kitaev \mbox{transformation~\cite{Bravyi:2002, Bravyi:2017}} to transform from a fermion to a qubit basis. The QCC ansatz operator $\hat{U}\l\bf{\tau}\r$ is specified according to the equation
\begin{equation}
\hat{U}\l\bf{\tau}\r = \prod_{k}^{n_{g}} \exp\l-\frac{\mathrm{i} \tau_{k}\hat{P}_{k}}{2}\r,
\end{equation}
where $\tau_{k}$ is a variational parameter, $n_g$ is the number of multi-qubit Pauli operators $\hat{P}_{k}$, defined as
\begin{equation}
\hat{P}_k = \bigotimes_{j}^{n_{q}} \hat{p}^{\l k\r}_{j}, \;\,\, \mathrm{for} \; \hat{p}^{\l k\r}_{j} \in \{\text{X, Y, Z, I\}},
\end{equation}
where $n_q$ is the number of qubits and X, Y, Z, and I are the Pauli matrices and a single-qubit identity operator, respectively. While the depth of the circuit rapidly increases as the size of the molecule increases, the QCC ansatz admits a low-depth quantum circuit compared to a widely used unitary coupled-cluster single and double ansatz. The details of the QCC circuit simulations can be found in Supplementary Note 1.

 We first consider an ansatz state that is a product state of $n_q$ arbitrary single-qubit states, following the method described in the work of Ryabinkin~et~al.~\cite{Ryabinkin:2020}. We evaluate the expectation value of the mean-field Hamiltonian with respect to the ansatz state and use VQE, simulated on a classical computer, to minimize the value. The variational parameters that result from the optimization correspond to the optimal wavefunction for the mean field. We next consider a QCC ansatz operator $\hat{U}\l\bf{\tau}\r$ applied to the previously determined, mean-field optimized state. Note that we aim to minimize the expectation value of the fragment Hamiltonian with respect to the ansatz operator parameters ${\bf \tau}$. We find which $\hat{P}_k$ to include in our ansatz by computing the derivative of the expectation value with respect to $\tau_k$, which can be computed in a straightforward way when ${\bf \tau} = {\bf 0}$. We remove $\hat{P}_k$ terms that have small derivative values. Applied to our example molecule H$_{10}$, we find that $\hat{P}_k$ of XY and YX have large, identical derivative values. Note that we carry out the computation of the derivatives on a classical computer.

We now investigate the performance of DMET with the QCC ansatz, which we denote as DMET-QCC, on a classical computer. Specifically, we consider the potential energy curve of the symmetric expansion (i.e., increasing the bond length $R$ while maintaining it for each pair of neighbouring atoms) of H$_{10}$. \revision{We choose 10 points $\l R=0.7, 0.85, 1.0, 1.1, 1.3, 1.6, 1.8, 2.0, 2.5, \text{and}~5.0~\si{\angstrom}\r$  along the potential energy curve and compare the total energies resulting from the two-qubit DMET-QCC ansatz with DMET-FCI, all calculated classically. The total energies per atom of the H$_{10}$ molecule are listed in Table~\ref{fig:dmet_energies} and we include the results obtained from other known methods, such as HF and full CI. The DMET-QCC results almost exactly reproduce the DMET-FCI results indicating that the VQE and QCC methods accurately represent the fragments.}

We next describe the simulation of our DMET method on a trapped-ion quantum computer. Instead of variationally optimizing the energies, then running through the DMET cycles, here we focus on the evaluation of the total energy from the quantum simulation for the classically pre-computed optimal parameters. Note that for the DMET cycles, we require $\langle XX\rangle$, $\langle YY \rangle$, $\langle ZZ \rangle$, $\langle XZ \rangle$, $\langle ZX\rangle$, $\langle Z_0\rangle$, $\langle Z_1\rangle$, $\langle X_0 \rangle$, and $\langle X_1 \rangle$ to be simulated, where the subscripts denote the qubit index. See Supplementary Note 2 for details on the Hamiltonian and the DMET energy expressions used in the experiments.

\subsection*{\revision{Compiling the circuits for the trapped ion hardware}}
The circuits are executed on IonQ's 11-qubit trapped-ion quantum computer, which is described in detail elsewhere~\cite{Wright_2019}. In the quantum computer, 15 ${}^{171}{\rm Yb}^+$ ions, aligned to form a linear crystal with spacing of about 4 $\mu$m, are suspended in a chip trap with a radial pseudopotential frequency of ${\sim}3.1$ MHz. We cool the crystal to its motional ground state and use the 11 ions in the middle as qubits, and initialize them to the $|0\rangle$ state. Two of these qubits are used in our experiment. Single- and two-qubit gate fidelities are nominally calibrated to greater than $99.5 \%$ and $96.5\%$, respectively.  Counter-propagating laser beams capable of illuminating individual ions are used to implement quantum gates, leveraging the ion--ion coupling mediated by the collective radial motional modes. We read out the quantum state by fluorescing the ions using a detection laser.

The native gates available on the quantum computer are $R_{\phi}(\pi/2) := \exp(-i \sigma_\phi \pi/4)$ and $\phi\phi'(\pi/2) := \exp(-i \sigma_{\phi} \sigma_{\phi'} \pi/4)$, where $\sigma_{\phi} = \cos(\phi)\text{X} + \sin(\phi)\text{Y}$.
We thus compile the DMET-QCC ansatz circuits discussed earlier using the native gates and optimize the circuits to reduce the number of quantum gates. Figure~\ref{fig:pipeline}(c) shows the pre- and post-optimizing compilation quantum circuits. See Methods section for further details. 

We extract the expectation values of the Pauli terms using the three compiled, optimized circuits shown in Figure~\ref{fig:pipeline}(c)(2)(i.). We use the XX and ZZ circuits for $\langle XX \rangle$ and $\langle ZZ \rangle$,  and the XZ circuit for $\langle XZ \rangle$ and $\langle ZX \rangle$, as the two pre-optimization circuits for $\langle XZ \rangle$ and $\langle ZX \rangle$ reduce to the same circuit upon optimizing  compilation. The single-qubit Pauli terms $\langle X_0 \rangle$, $\langle X_1 \rangle$, $\langle Z_0 \rangle$, and $\langle Z_1 \rangle$, where $B_i$ denotes the $i$-th qubit's expectation value in the $B$ basis, are computed using all of the statistics available from the three circuit executions. For instance, for $\langle X_0 \rangle$, we use the results from both $\langle XX \rangle$ and $\langle XZ \rangle$. Likewise approaches are used for $\langle X_1 \rangle$, $\langle Z_0 \rangle$, and $\langle Z_1 \rangle$.

\subsection*{\revision{Experimental results}}
\begin{figure*}[hbtp]
\centering
\includegraphics[width=\textwidth]{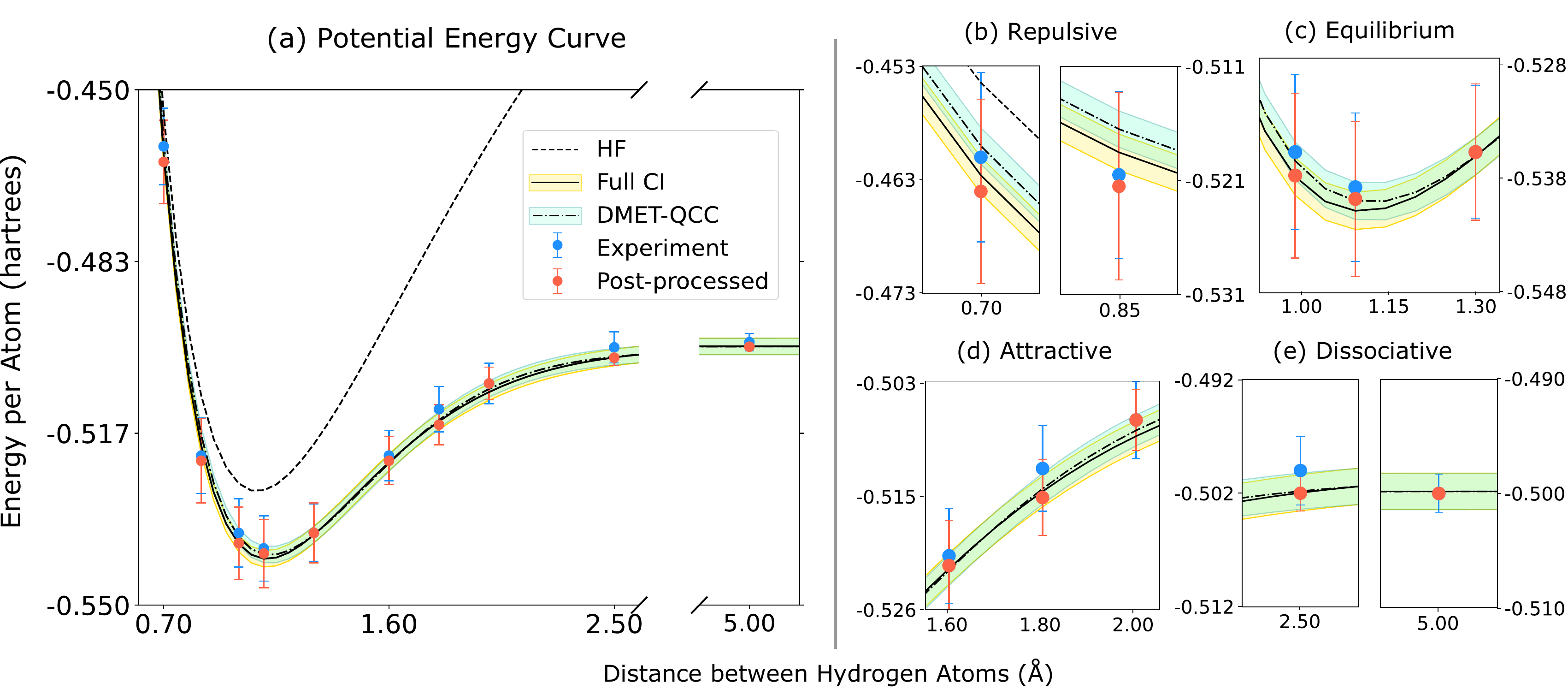}
\caption{
The potential energy curve of the symmetric expansion of the ring of 10 hydrogen atoms (H$_{10}$). For reference, we include classically simulated curves of HF \revision{(dotted black line)}, full CI \revision{(black line)}, and DMET-QCC \revision{(dashed black line)}, the last two surrounded by yellow and blue regions indicating chemical accuracy with respect to their values, and their overlap indicated in green. The energies obtained by experiment are shown in blue \revision{circles}, and the post-processed energies are shown in red \revision{circles}. The respective error bars for each are calculated using bootstrapping. (a) The full potential energy curve. \mbox{(b)--(e) A magnified} view of the distinct energy regimes of the curve.} 
\label{fig:h10_pes}
\end{figure*}

The total energies obtained from the quantum computer (points plotted in blue), along with the classically computed potential energies (lines), are shown in Fig.~\ref{fig:h10_pes}. We calculate the values of the error bars using the bootstrapping method described in \revision{the Methods section}. The experimentally determined energies nearly coincide with the full CI energies (black line) and DMET-QCC reference values, and agree within the margin of error.

To improve our experimental results, we perform classical post-processing using McWeeney's density matrix purification technique~\cite{Truflandier:2016} applied to the reduced density matrices (RDM). Its effectiveness in improving energy estimation has been shown in~\cite{Arute:2020} for uncorrelated one-particle reduced density matrices (1-RDM), and in~\cite{McCaskey:2019} for two-particle reduced density matrices (2-RDM) of correlated two-electron systems. As in the latter case, ours is a two-electron system whose 2-RDM is the full density matrix $P_{pqrs}$. \revision{Only because of this aspect,} we are able to purify our experiment iteratively using the following procedure: \mbox{$P^{\text{new}}_{pqrs}=3(P^{\text{old}}_{pqrs})^{2}-2(P^{\text{old}}_{pqrs})^{3}$} until the convergence criterion of $\text{Tr}(P^{2}_{pqrs}-P_{pqrs}) < 1.0 \times 10^{-2}$ has been met, see Methods section for details.

\revision{We emphasize that this purification only works when applied to the 2-RDM for our system because our fragment only has two electrons -- thus the 2-RDM happens to be the full density matrix, allowing idempotency to be imposed.} To extend this method to larger correlated systems, we would \revision{need to} purify our 2-RDMs using the more general $n$-representability conditions~\cite{rubin2018,Lanssens2018,Bach2015}. \revision{This requires further study and is outside the scope of this work}. Obtaining $P_{pqrs}$ requires us to run an additional circuit, YY, on the trapped-ion quantum computer. The corresponding circuit is shown in Fig.~\ref{fig:pipeline}(c)(2)(ii.) comprising three single-qubit and one two-qubit native gate operations; see Methods section for details on the optimized compilation.

Table~\ref{fig:dmet_energies} and the  points plotted in red in Fig.~\ref{fig:h10_pes} show the post-processed total energies for all points, along with their respective errors. All points are in agreement with the simulated DMET-QCC values, and, with the exception of the repulsive wall and the point $R=1.8~\si{\angstrom}$, all values are within chemical accuracy. Again excluding the repulsive wall, the purification method improves the accuracy of the energy estimate for all points along the potential energy curve, and brings the point $R=2.5~\si{\angstrom}$ within chemical accuracy of DMET-QCC and full CI. This post-processing is enabled by the high quality of the two-qubit gates of the hardware; see Supplementary Note 3 for further details.

\subsection*{\revision{Discussion}}
\revision{The level of accuracy in our results is due to the ability of the decomposition method to well approximate most of the curve, the success of the VQE algorithm and QCC circuit ansatz to express the fragments, } and the small systematic errors of the hardware used in this experiment. We note that the error bars are quite large, which is a result of the limited sampling capacity. The achievable precision will increase as the availability of quantum resources grows. \revision{Simulating a molecule of this size was only possible through using problem decomposition to create fragments that would fit the hardware. This demonstrates how we can use classical methods to extend the applicability of quantum hardware that is limited in size, and scale the correlation captured in our fragments as hardware develops.}

In order to capture the nature of the entanglement in our simulation, which is the characteristic phenomenon of quantum simulation, we also investigate two types of quantum entanglement. The Hilbert space of the two-qubit system is split into two subsystems: the first and second qubits, and the fragment and the bath orbitals. In both cases, the experimental results are very close to theoretical predictions and clearly show that the states we observe in the present experiment are entangled. See Supplementary Note 4 for more details.

\revision{\section*{Conclusion}}

\revision{In this work, we have experimentally demonstrated an end-to-end pipeline using a PD technique to reduce the size of an electronic structure simulation, making a 20 qubit problem realizable on a trapped-ion NISQ device. Our method involves} combining DMET and VQE with a QCC ansatz to compress the quantum simulation circuit for the electronic structure calculation, and circuit optimization techniques that target trapped-ion quantum computers. A density matrix purification method is then applied to mitigate residual errors. \revision{All of these steps successfully construct the potential energy curve of the ring of 10 hydrogen atoms, and serves as a proof of concept for the usefulness of problem decomposition to accurately and efficiently represent large molecules.} Our results are in agreement with full CI, and almost all points are within chemical accuracy of their DMET-QCC simulated values. \revision{This} demonstrates that the approach employed herein can describe bond-breaking and bond-forming in molecular systems.

Computer-aided molecular design in both materials science and the life sciences is key to our attaining a sustainable future through the accurate prediction of chemical reactions, such as the catalytic reaction of organometallic compounds in advanced materials innovation, enzymatic reactions in the life sciences, and the electrochemical reactions implemented in next-generation batteries. In a previous work by some authors of the present paper~\cite{Verma:2020}, it is shown that a quantum simulation of a typical, industrially relevant organometallic compound would require over 2000 qubits. Selecting the appropriate PD technique could result in the reduction of qubit requirements significantly. A reduction by a factor of five for industrially relevant systems is shown in another work~\cite{Yamazaki:2018}. This sort of reduction could place these simulations  within the reach of near-future NISQ devices that have hundreds of qubits. The precise reduction in the amount of required resources can be both algorithm dependent and application dependent. For example, if we were to use the DMET algorithm as the PD technique, and if our target system were not symmetric, we would need to run quantum simulations for each fragment that the DMET algorithm would create. If we were to use a larger fragment size for greater accuracy, the quantum resource required to simulate the fragment would also become greater, but would always remain smaller than or equal in size to the case where the entire system were simulated without using any PD. 

Our present work experimentally demonstrates the potential of problem decomposition methods as a pre-processing step to reduce the quantum resources required in simulating molecular systems while preserving accuracy. \revision{We were able treat all electrons in the system, and include correlations as necessary to produce accurate results using subproblems that fit on hardware.} \revision{This additional classical component allows us to apply results from quantum devices to a larger family of problems, which we believe will be a valuable tool for using quantum computers to enhance molecular design platforms.}

\revision{\section*{Methods}}
\subsection*{The System Hamiltonian}
The second-quantized electronic Hamiltonian can be written as
\begin{align}
\hat{H} = \sum_{pq}h_{pq} \hat{a}^{\dag}_p a_q + \frac{1}{2} \sum_{pqrs} \big(pq|rs\big) \hat{a}^{\dag}_p\hat{a}^{\dag}_s\hat{a}_r\hat{a}_q\, ,
\end{align}
where \(p\), \(q\), \(r\), and \(s\) are distributed over all spin-orbitals, and \(a^{\dag}_p\) and \(a_q\) are the corresponding creation and annihilation operators. The terms \(h_{pq}\) and $\big(pq|rs\big)$ are the one- and two-electron integrals (in chemists' notation), respectively. The evaluation of these integrals, as well as the Hartree--Fock (HF) and full CI calculations, are carried out using PySCF~\cite{Sun:2018}. The minimal basis set MINAO~\cite{Knizia:2013} is used in our calculation. 

The electronic Hamiltonian is then transformed into the qubitized form by using symmetry-conserving Bravyi--Kitaev transformation (scBK)~\cite{Bravyi:2002, Bravyi:2017}.
The qubitized Hamiltonian can be written as 
\begin{equation}
\hat{H} = \sum_p h_p^{\alpha} \sigma_p^{\alpha} + \sum_{pq}h_{pq}\sigma_{p}^{\alpha}\sigma_{q}^{\beta} + \sum_{pqr}h_{pqr}\sigma_{p}^{\alpha}\sigma_{q}^{\beta}\sigma_{r}^{\gamma} + \ldots,
\end{equation}
where \(p\), \(q\), \(r\),\(\dots\) are distributed over all qubits, and \(\sigma_p^{\alpha} \in \{\sigma_x, \sigma_y, \sigma_z, I\}\) acts on qubit~\(p\). The transformation of the electronic Hamiltonian into a qubit basis is performed using OpenFermion~\cite{McClean:2017}.

\subsection*{Density Matrix Embedding Theory}

Let us consider a molecular system divided into fragments, in which a small fragment A with $N^{\text{A}}$ states is surrounded by a large bath B with $N^{\text{B}}$ states. If the wavefunction of the entire system $|\Psi\rangle$ is known, Schmidt decomposition~\cite{Peschel:2012}  can be applied, and we may write
\begin{equation}
\begin{split}
|\Psi\rangle &= \sum_{i}^{N^{\text{A}}}\sum_{j}^{N^{\text{B}}}\psi_{ij}|\alpha_{i}\rangle|\beta_{j}\rangle \\
&= \sum_{i}^{N^{\text{A}}}|\alpha_{i}\rangle\Bigg(\sum_{j}^{N^{\text{B}}}\psi_{ij}|\beta_{j}\rangle\Bigg) \\
&= \sum_{i}|\alpha_{i}\rangle|\Tilde{\chi_{i}}\rangle \\
&= \sum_{ii'}\psi_{ii'}|\alpha_{i}\rangle|\chi_{i'}\rangle\,,
\end{split}
\label{eq: Schmidt decomp}
\end{equation}
where   
\begin{equation}
    |\Tilde{\chi_{i}}\rangle = \sum_{j}\psi_{ij}|\beta_{j}\rangle,
\end{equation}
$|\alpha_{i}\rangle$ and $|\beta_{i}\rangle$ denote particular many-body bases, and $|\chi_{i}\rangle$ is the orthogonal set of $|\Tilde{\chi_{i}}\rangle$. The states $|\chi_{i'}\rangle$ are the states of bath B; however, the number of states coincides with that of fragment A. This shows that, regardless of the size of bath B, only $N^{\text{A}}$ states can be entangled with the fragment. This will reduce the size of the problem drastically for large-sized systems. 

The exact wavefunction of the entire system $|\Psi\rangle$ is the eigenfunction of the Hamiltonian for the entire system $H$. The Hamiltonian $H'$ of fragment A embedded in bath B can be defined using a projection operator $P$, that is,
\begin{equation}
    H' = PHP\,,
\end{equation}
where
\begin{equation}
    P = \sum_{ii'}|\alpha_{i}\rangle|\chi_{i'}\rangle\langle\alpha_{i}|\langle\chi_{i'}|.
\end{equation}

It is now evident that the electronic structure of the entire system can be described exactly by that of the fragments and their surrounding baths. The electronic structure calculation of the entire system can be solved using this smaller problem. However, the exact wavefunction of the molecular system is usually not known a priori; thus, introduction of an approximation is necessary. The wavefunction of the entire system obtained by low-level, mean-field theory, such as the HF calculation, would be a straightforward approximation of the exact wavefunction of the entire system. Using a low-level wavefunction to construct a bath and solve the reduced problem employing a high-level theory is the principal idea behind the density matrix embedding theory (DMET). 

The orbitals of the entire system are transformed into unentangled occupied orbitals, unentangled virtual orbitals, local fragment orbitals, and bath orbitals. The orbital space of the entire system is then greatly reduced, as only the local fragment and bath orbitals are employed for each high-level DMET fragment calculation.

Practically, we need to optimize the embedding of a bath. In DMET, a high-level calculation for each fragment is carried out individually until self-consistency has been attained according to a certain criterion: the sum of the 1-RDM of all of the fragments agrees with that of the low-level one for the entire system. The DMET energy is calculated using the 1-RDM and 2-RDM. The DMET algorithm used in this work, the single-shot algorithm~\cite{Wouters:2016}, can be described as follows: 
\begin{enumerate}
    \item Calculate the wavefunction of the entire molecular system using a low-level method and then localize the orbitals to fragment the molecule.
    \item Construct the bath orbitals so as to include the surrounding environmental effect.
    \item Construct the Hamiltonian of a fragment (including environmental effect) and calculate the wavefunction using a method based on a high-level theory.
    \item Calculate the fragment energy and the number of electrons for each fragment with the 1- and 2-RDMs from the wavefunction obtained in Step 3. 
    \item Repeat steps 2--4 for each fragment and obtain the total energy and the high-level 1-RDM of the entire molecular system.
    \item Repeat steps 2--5 until the sum of the number of electrons in the fragments agrees with the number of electrons for the entire system.
\end{enumerate}

The initial step of a DMET calculation is to perform a mean-field HF calculation for the entire molecule. The localized orbitals are obtained by localizing the canonical orbitals from the HF calculation to determine how to fragment the molecules. The Meta-L{\"o}wdin localization scheme~\cite{Sun:2014} is used in this work.

After the molecule is divided, the DMET cycle (steps 2--6) is initiated. The first step in the cycle is to obtain the bath orbitals $\hat{a}_{r}^\dagger$ (the active orbitals in the electronic structure calculation used for fragments to describe  environmental effects), and the environment density matrix of fragment A,  $D^{\text{env,A}}$, is calculated as follows: 
\begin{equation}
\label{eq:env_dm}
D^{\text{env,A}}_{pq}=\sum_{r \in \text{env}}C_{pr}C^{\dagger}_{rq},
\end{equation}
where $C$ represents the molecular orbital coefficients obtained from the mean-field calculation of the entire molecule. Now, the embedding Hamiltonian $H^{\text{emb,A}}$ can be constructed (step 3). It can be defined as 
\begin{equation}
\label{eq:embed_hamiltonian}
\begin{split}
H^{\text{emb,A}}&=\sum_{pq}^{L^{\text{A}}+L^{\text{B}}}[h_{pq}+ \sum_{rs}^{L}[\big(pq|rs\big) -\big(ps|rq\big)]D_{rs}^{\text{env}}]\hat{a}_{p}^{\dagger} \hat{a}_{q} \\
&-\delta\mu\sum_{p \in \text{A}}\hat{a}_{p}^{\dagger}\hat{a}_{p} +\sum_{pqrs}^{L^\text{{A}}}\big(pq|rs\big)\hat{a}_{p}^{\dagger} \hat{a}_{r}^{\dagger} \hat{a}_{s} \hat{a}_{q}\,,    
\end{split}
\end{equation}
where the $h_{pq}$ are the one-electron integrals, the $\big( pq|rs \big)$ are the two-electron integrals in chemists'
notation, $L^\text{{A}}$ is the number of orbitals in the fragment, $L^\text{{B}}$ is the number of bath orbitals, $L$ is the number of orbitals in the entire molecule, and $p$, $q$, $r$, and $s$ are general orbital indices. We introduce the chemical potential $\delta\mu$, which is optimized in the DMET cycle. Once the Hamiltonian $H^{\text{emb,A}}$ has been obtained, the electronic structure calculation is performed for fragment A. We employ VQE with the QCC ansatz in this work. Following these calculations, the algorithm constructs the one- and two-particle density matrices, $D^\text{{A}}$ and $P^\text{{A}}$, respectively, from the QCC wavefunction $\Psi^\text{{A}}$. The fragment energy $E^\text{{A}}$ is calculated (step 4) from the RDMs as 
\begin{equation}
\label{eq:frag_energy}
\begin{split}
E^\text{{A}}&=\sum_{p \in \text{A}}\Bigg(\sum_{q}^{L^\text{{A}}+L^\text{{B}}} \\
&\quad \Big(h_{pq}+\frac{\sum_{rs}^{L}[\big(pq|rs\big)-\big(ps|rq\big)]D_{rs}^{\text{env}}}{2}\Big)D_{qp}^\text{{A}} \\
&+\frac{1}{2}\sum_{qrs}^{L^\text{{A}}+L^\text{{B}}}\big(pq|rs\big)P_{qp|sr}^\text{{A}}\Bigg).
\end{split}
\end{equation}
Note that only the elements with fragment orbital indices ($p \in$ A) are used for calculating the fragment energy. Here, the 1- and 2-RDMs are defined as
\begin{equation}
\label{eq:one_rdm}
D_{qp}^\text{{A}}=\big \langle \hat{a}_{p}^{\dagger} \hat{a}_{q} \big \rangle
\end{equation}
and
\begin{equation}
\label{eq:two_rdm}
P_{qp|sr}^\text{{A}}=\big \langle \hat{a}_{p}^{\dagger} \hat{a}_{r}^{\dagger} \hat{a}_{s} \hat{a}_{q} \big \rangle,
\end{equation}
respectively. The number of electrons $N^\text{A}$ in fragment A are calculated (step 4) as
\begin{equation}
\label{eq:frag_ele}
N^\text{{A}}=\sum_{p \in \text{A}}D_{pp}^\text{{A}}\,.
\end{equation}
The DMET energy is calculated by summing the fragment energy for each fragment (step 5), which is obtained according to the equation
\begin{equation}
\label{eq:dmet_ene}
E^{\text{DMET}}=\sum_{A}E^{A}+E^{\text{nuc}},
\end{equation}
where $E^{\text{nuc}}$ is the nuclear repulsion  energy.
The DMET cycle (steps 2--6) iterates until the number of electrons in the DMET calculation given by 
\begin{equation}
\label{eq:dmet_ele}
N^{\text{frag}}=\sum_{A}N^{A}
\end{equation}
converges to the total number of electrons in the molecule $N^{\text{tot}}$. Convergence is achieved by updating the chemical potential $\delta\mu$ according to the equation
\begin{equation}
\label{eq:chemical_potential}
\delta\mu=a(N^{\text{frag}}-N^{\text{tot}}),
\end{equation}
where $a$ is positive number.

For the DMET calculation for the H$_{10}$ ring, the high symmetry of the molecular structure allows us to reduce the calculation further. We calculate only a single fragment with the assumption that all of the fragments have both the same energy and number of electrons. The DMET total energy with only one fragment multiplied by 10 (the number of hydrogen atoms) coincides with the energy when using all fragments; thus, we treat only one fragment in the DMET calculation and simply multiply the energy and the number of electrons by 10.

\subsection*{The Qubit Coupled-Cluster Method}
An accurate and affordable description of the correlated wave function $\Ket{\bf{\Psi}}$ required to evaluate the energy of fragment~A according to Eq.~(\ref{eq:frag_energy}) is achieved using the qubit coupled-cluster (QCC) method~\cite{Ryabinkin:2018}.
Within the QCC approach, a mean-field wave function $\Ket{\bf{\Omega}}$ is determined and subsequently utilized in a heuristic~\cite{Ryabinkin:2020} to construct a unitary operator ansatz $\hat{U}\left(\bf{\tau}\right)$.
This operator recovers the missing electron correlation for the mean-field state $\Ket{\bf{\Omega}}$ and results in the QCC wave function according to the equation
\begin{equation} \label{eq:qcc1}
  \Ket{\mathbf{\Psi}\l\bf{\tau},\mathbf{\Omega}\r} = \hat{U}\l\bf{\tau}\r \Ket{\mathbf{\Omega}}.
\end{equation}
A parameterized mean-field wave function is defined as a tensor product of $n_{q}$ single-qubit states and can be expressed as
\begin{equation} \label{eq:qcc2}
  \Ket{\mathbf{\Omega}\l\mathbf{\Gamma}\r} = \bigotimes_{j}^{n_{q}} \Ket{\bf{\omega}\l\theta_{j}, \phi_{j}\r},
\end{equation}
where $\bf{\Gamma} = \{\theta_{j}\}_{j=1}^{n_{q}} \cup \{\phi_{j}\}_{j=1}^{n_{q}}$ is the set of $2n_q$ mean-field parameters.
Each single-qubit state is then represented in the qubit computational basis as
\begin{equation} \label{eq:qcc3}
  \Ket{\bf{\omega}\l\theta_{j}, \phi_{j}\r} = \cos\l\frac{\theta_{j}}{2}\r\Ket{\mathbf{0}_{j}} +  \e^{\mathrm{i} \phi_{j}}\sin\l\frac{\theta_{j}}{2}\r\Ket{\mathbf{1}_{j}}, 
\end{equation}
and is characterized by the Bloch sphere polar and azimuthal angles $\theta_{j}$ and $\phi_{j}$, respectively \cite{Ryabinkin:2018}.
The mean-field energy functional is given by
\begin{equation} \label{eq:qcc4}
  E^{\text{MF}}\l\mathbf{\Gamma}\r = \Bra{\mathbf{\Omega}\l\mathbf{\Gamma}\r} \hat{H} \Ket{\mathbf{\Omega}\l\mathbf{\Gamma}\r}.
\end{equation}
Minimization of Eq.~(\ref{eq:qcc4}) with respect to $\bf{\Gamma}$ gives the ground-state, mean-field energy $E^{\text{MF}}$ and the corresponding optimal set of parameters $\bf{\Gamma}^{\text{opt}}$.
The QCC unitary operator ansatz $\hat{U}\l\bf{\tau}\r$ takes the form
\begin{equation} \label{eq:qcc5}
\hat{U}\l\bf{\tau}\r = \prod_{k}^{n_{g}} \exp\l-\frac{\mathrm{i} \tau_{k}\hat{P}_{k}}{2}\r,
\end{equation}
where $\hat{P}_{k}$ is a multi-qubit Pauli operator defined as
\begin{equation} \label{eq:qcc6}
\hat{P}_k = \bigotimes_{j}^{n_{q}} \hat{p}^{\l k\r}_{j} \; \mathrm{for} \; \hat{p}^{\l k\r}_{j} \in \{\text{X, Y, Z, I}\},
\end{equation}
$\tau_{k}$ is a variational parameter, and $n_g$ is the number of Pauli operators $\hat{P}_{k}$ included in the ansatz.
The QCC energy functional is defined as
\begin{equation} \label{eq:qcc7}
  E^{\text{QCC}}\l\bf{\tau}, \mathbf{\Gamma}\r = \Bra{\mathbf{\Omega}\l\mathbf{\Gamma}\r} \hat{U}^{\dagger}\l\bf{\tau}\r \hat{H} \hat{U}\l\bf{\tau}\r\Ket{\mathbf{\Omega}\l\mathbf{\Gamma}\r},
\end{equation}
where $\bf{\tau} = \{\tau_{k}\}_{k=1}^{n_g}$ is the set of $n_{g}$ variational parameters.
The heuristic screening procedure~\cite{Ryabinkin:2020} is utilized to construct the set of operators $\{\hat{P}_{k}\}_{k=1}^{n_g}$ appearing in Eq.~(\ref{eq:qcc5}).
This heuristic approach~\cite{Ryabinkin:2018} relies on the gradient of Eq.~(\ref{eq:qcc7}) with respect to $\tau_{k}$ evaluated using the optimal mean-field wave function $\l\text{i.e.,} \; \bf{\Gamma} = \bf{\Gamma}^{\text{opt}} \; \text{and} \; \bf{\tau} = \bf{0}\r$, the form of which is
\begin{equation} \label{eq:qcc8}
  \left.\frac{\d E^{\text{QCC}}}{\d \tau_{k}}\right\vert_{\bf{\tau} = \bf{0}} = \frac{i}{2} \Bra{\mathbf{\Omega}\l\mathbf{\Gamma}^{\text{opt}}\r} [\hat{P}_{k}, \hat{H}] \Ket{\mathbf{\Omega}\l\mathbf{\Gamma}^{\text{opt}}\r},
\end{equation}
where $[\hat{P}_{k}, \hat{H}] = \hat{P}_{k} \hat{H} - \hat{H} \hat{P}_{k}$.
Equation~(\ref{eq:qcc8}) is quantified for a representative $\hat{P}_{k}$ from each group of electron correlation generators contained in the direct interaction set (DIS)~\cite{Ryabinkin:2020}.
The $n_g$ representative generators from the DIS with the largest energy gradient magnitudes are utilized in Eq.~(\ref{eq:qcc5}).
Minimization of the energy functional given by Eq.~(\ref{eq:qcc7}) with respect to $\bf{\tau}$ gives the QCC ground state energy $E^{\text{QCC}}$ and the corresponding optimal set of parameters $\bf{\tau}^{\text{opt}}$.
We note that, in the present work, the optimal mean-field parameter set $\bf{\Gamma}^{\text{opt}}$ is not relaxed during the minimization of Eq.~(\ref{eq:qcc7}).
The  parameter set $\bf{\mathcal{T}}^{\text{opt}}$ of size $2 n_q + n_g$ that specifies the ground state QCC correlated wave function $\Ket{\bf{\Psi}\l\bf{\mathcal{T}}^{\text{opt}}\r}$ is formed as the union of the separately optimized mean-field and QCC parameter sets: $\bf{\mathcal{T}}^{\text{opt}} = \bf{\Gamma}^{\text{opt}} \cup \bf{\tau}^{\text{opt}}$.

The DMET-QCC framework is applied to compute the energy of each point along the H$_{10}$ potential energy curve (see Table~\ref{fig:dmet_energies} and Fig.~\ref{fig:h10_pes}).
The pre-optimized quantum circuits utilized for the trapped ion hardware experiments are constructed with $\bf{\mathcal{T}}^{\text{opt}}$ obtained from classical DMET-QCC  simulations of the embedding Hamiltonian for fragment A (see Eq.~(\ref{eq:embed_hamiltonian})).
At each point $R$ along the potential energy curve, $\hat{H}^{\text{emb,A}}\l R\r$ is a two-qubit operator $\l n_q = 2\r$ after scBK encoding and takes the form
\begin{equation} \label{eq:qcc9}
  \begin{aligned}
    \hat{H}^{\text{emb,A}}\l R\r & = a\l R\r + b\l R\r \text{X}_{0}\text{X}_{1} + c\l R\r \text{Z}_{0}\text{Z}_{1} \\
                                       & + d\l R\r [\text{X}_{0} + \text{X}_{1}] + e\l R\r [\text{Z}_{0} + \text{Z}_{1}]  \\
                                       & + f\l R\r [\text{X}_{0}\text{Z}_{1} + \text{Z}_{0}\text{X}_{1}].
  \end{aligned}
\end{equation}
The inter-atom, spacing-dependent expansion coefficients of $\hat{H}^{\text{emb,A}}\l R\r$ in Eq.~(\ref{eq:qcc9}) are provided in Supplementary Note 1 for each point along the H$_{10}$ potential energy curve. ProjectQ~\cite{Steiger:2018} is employed to perform simulation on classical hardware. Further details of the DMET-QCC circuits and simulations can also found in Supplementary Note 1.

\subsection*{Density Matrix Purification}
We perform density matrix purification based on McWeeny's purification scheme~\cite{Truflandier:2016}. This iterative method purifies the 2-RDM $P_{pqrs}$ according to 
\begin{equation}
P^{\text{new}}_{pqrs}=3(P^{\text{old}}_{pqrs})^{2}-2(P^{\text{old}}_{pqrs})^{3}.    
\end{equation}
The iteration is continued until the convergence criterion
\begin{equation}
\text{Tr}(P^{2}_{pqrs}-P_{pqrs}) < \epsilon
\end{equation}
is met. In this work, $\epsilon=1.0 \times 10^{-2}$ is used. Note that the change in the total energy is smaller than a millihartree when we changed the criterion to $1.0 \times 10^{-7}$. The tensor multiplication of $P$ is defined as 
\begin{equation}
A_{pqrs}=B_{pquv}*C_{uvrs},    
\end{equation}
where the Einstein summation is implied.

Note that this purification technique can only be applied to two-electron systems~\cite{McCaskey:2019, Arute:2020}. Although we here consider a 10-electron system, the RDM in our current work can be purified, as the fragment calculation for DMET involves a two-electron system. \revision{More analysis of the purification method applied to the experimental results can be found in Supplementary Note 4.}

\subsection*{The Bootstrap Method}
The total energies and their statistical errors are calculated using an empirical bootstrapping method. We  follow the procedure described in a previous work~\cite{Nam:2019}. We start from the state preparation and measurement (SPAM)-corrected histograms for each Pauli operator (XX, YY, ZZ, XZ, and ZX), and construct a distribution of total energies.

The mean and the standard deviation ($\sigma$) are computed from the distribution of energies. The procedure of the bootstrapping method is as follows.

\begin{enumerate}
    \item Draw a random bootstrap sample of the same size as the original dataset with replacement of the data.
    \item Construct a new histogram based on step 1.
    \item Compute the expectation value of the Pauli terms using the new histogram.
    \item Repeat steps 1--3 for each Pauli term and obtain all expectation values needed to construct RDMs.
    \item Construct 1- and 2-RDMs.
    \item Calculate the total energy.
    \item Repeat steps 1--6 10,000 times and obtain a distribution of total energies.
    \item Calculate the mean and the standard deviation from the distribution of energies constructed in step~7. 
\end{enumerate}

We follow this procedure to construct a histogram of possible measurements consistent with the empirical data. The calculated value of $\sigma$ is represented using error bars. The mean of this distribution is the measured energy, and the 1$\sigma$ error estimate is its standard deviation. The difference from the previous work~\cite{Nam:2019} is the construction of the RDMs (step 5), which is required for DMET energy calculation. 
Note that when we perform density matrix purification, we purify the 2-RDM between steps 5 and 6. 

\section*{Circuit Optimization and Compiling}

Our DMET-QCC ansatz circuits for ${\rm H}_{10}$ are written as a standard gate set that consists of \mbox{controlled-NOT,} $R_x(\theta) := \exp(-i\sigma_x\theta/2)$, $R_y := \exp(-i\sigma_y\theta/2)$, and $R_z := \exp(-i\sigma_z\theta/2)$ gates. To implement these circuits on a trapped-ion quantum computer, we transpile them so that the output circuits are encoded in the trapped-ion gate set. This set consists of $\phi\phi' := \exp(-i\sigma_{\phi} \sigma_{\phi'}\pi/4)$ and $\phi := \exp(-i\sigma_{\phi}\pi/4)$ gates, where $\sigma_{\phi} := \cos(\phi)\sigma_x + \sin(\phi)\sigma_y$. Further, we keep only three decimal digits after the decimal point in all our gate parameter specifications, commensurate with the machine's level of precision.

\begin{figure*}
\centering
\includegraphics[width=0.5\textwidth]{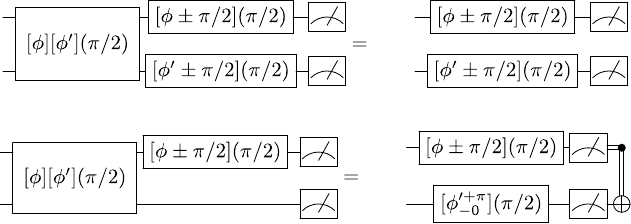}
\caption{Circuit optimization. The rules used to optimize our DMET-QCC ansatz circuits.}
\label{fig:circ_opt} 
\end{figure*}

We combine the circuit compilation and optimization techniques reported in~\cite{Nam:2019,Maslov_2017,Nam_2018} with the circuit optimization technique for the trapped-ion gate set shown in Fig.~\ref{fig:circ_opt} to obtain the final, optimized circuits, amenable to implementation on a trapped-ion quantum computer. The pre- and post-optimization circuits are represented in Fig.~\ref{fig:pipeline}(c). Their gate parameters appear in Tables~\ref{tab:param-pre} and \ref{tab:param-post}.

Specifically, to perform the classical post-processing step, we obtain the $\langle YY\rangle$ expectation value from the trapped-ion quantum computer.
The pre-optimization circuit is the same as that used to calculate any other expectation value,
except the measurement basis transformation operations
$M_0$ and $M_1$ in Fig.~\ref{fig:pipeline}(c)(2)(i.) are $\sgate^\dagger \hgate$.
The post-optimization circuit is shown in Fig.~\ref{fig:pipeline}(c)(2)(ii.).
Note that we use a real degree of freedom in the two-qubit gate,
that is, $\phi\phi'(\theta) := \exp(-i\sigma_{\phi} \sigma_{\phi'}\theta/2)$,
as has been used elsewhere for efficient and high-fidelity simulations~\cite{Shehab_2019,Nam:2019}. The gate parameter values of $\theta$ for each point along the potential energy curve are shown in Table~\ref{tab:param-post2}.

\begin{table*}[hbtp]
\centering
\caption{\revision{Numerical and experimental values of the total energies per atom of the ring of 10 hydrogen atoms (H$_{10}$)}. The calculated total energies per atom (in hartrees), using the HF, full CI, \revision{DMET-FCI,} and DMET-QCC methods, for the H$_{10}$ molecule along the potential energy curve $\l R=0.7, 0.85, 1.0, 1.1, 1.3, 1.6, 1.8, 2.0, 2.5, \text{and}~5.0~\si{\angstrom}\r$ of the symmetric expansion. For DMET-QCC, we report the theoretical (T), experimental (E), and post-processed (P) results. For the experimental and the post-processed results, we report the standard deviation of the energies, calculated using bootstrapping.}
\vspace{0.6em}
\footnotesize{
\begin{tabular}{ c | c | c | c | c | c | c}
$R\l\si{\angstrom}\r$ & HF & FCI & \revision{DMET-FCI} & DMET-QCC (T) & DMET-QCC (E) & DMET-QCC (P) \\
\hline
0.70 & $-$0.454468 & $-$0.462588 & \revision{$-$0.460015} & $-$0.460015 & $-$0.461 $\pm$ \, 0.007 & $-$0.464 $\pm$ \, 0.008 \\
0.85 & $-$0.509296 & $-$0.519025 & \revision{$-$0.516992} & $-$0.516991 & $-$0.521 $\pm$  0.007 & $-$0.522 $\pm$ 0.008\\
1.00 & $-$0.526412 & $-$0.538093 & \revision{$-$0.536753} & $-$0.536753 & $-$0.536 $\pm$ \, 0.007 & $-$0.538 $\pm$ \, 0.007 \\
1.10 & $-$0.527728 & $-$0.541007 & \revision{$-$0.540160} & $-$0.540160 & $-$0.539 $\pm$ \, 0.006 & $-$0.540 $\pm$ \, 0.007 \\
1.30 & $-$0.518798 & $-$0.536375 & \revision{$-$0.536353} & $-$0.536354 & $-$0.536 $\pm$ \, 0.006 & $-$0.536 $\pm$ \, 0.006 \\
1.60 & $-$0.494352 & $-$0.522320 & \revision{$-$0.522484} & $-$0.522484 & $-$0.521 $\pm$ \, 0.005 & $-$0.522 $\pm$ \, 0.004 \\
1.80 & $-$0.476771 & $-$0.514366 & \revision{$-$0.514092} & $-$0.514093 & $-$0.512 $\pm$  0.004 & $-$0.515 $\pm$  0.004\\
2.00 & $-$0.460103 & $-$0.508668 & \revision{$-$0.508073} & $-$0.508074 & $-$0.507 $\pm$  0.004 & $-$0.507 $\pm$  0.003\\
2.50 & $-$0.425756 & $-$0.501961 & \revision{$-$0.501820} & $-$0.501822 & $-$0.500 $\pm$  0.003 & $-$0.502 $\pm$  0.001\\
5.00 & $-$0.368206 & $-$0.499810 & \revision{$-$0.499826} & $-$0.499826 & $-$0.499 $\pm$  0.002 & $-$0.49987 $\pm$  0.00008\\
\hline
\end{tabular}
}
\label{fig:dmet_energies}
\end{table*}

\begin{table*}[hbtp]
\centering
\caption{ Pre-optimized parameter specification. Parameters of the gates that appear in the pre-optimization circuits in Fig.~\ref{fig:pipeline}(c)(i.). All parameters are in radians.}
\vspace{0.6em}
\footnotesize{
\begin{tabular}{ c | c | c | c | c | c | c | c }
$R\l\si{\angstrom}\r$ & $\rzgate^{(0,0)}$ & $\rygate^{(0,1)}$ & $\rzgate^{(1,0)}$ & $\rygate^{(1,1)}$ & $\rxgate^{(1,2)}$ & $\rzgate^{(1,3)}$ & $\rxgate^{(1,4)}$ \\
\hline
0.7 & 0.437 & 3.142 & 0.395 & 3.142 & 1.571 & 0.074 & 10.996 \\
0.85 & 0.000 & 3.142 & 0.000 & 3.142 & 1.571 & 0.106 & 10.996 \\
1.0 & 4.618 & 3.142 & 3.617 & 3.142 & 1.571 & 0.145 & 10.996 \\
1.1 & 5.383 & 3.142 & 0.097 & 3.142 & 1.571 & 0.176 & 10.996 \\
1.3 & 3.577 & 3.142 & 0.103 & 3.142 & 1.571 & 0.249 & 10.996 \\
1.6 & 3.714 & 3.142 & 1.481 & 3.142 & 1.571 & 0.395 & 10.996 \\
1.8 & 0.000 & 3.142 & 0.000 & 3.142 & 1.571 & 0.520 & 10.996 \\
2.0 & 0.000 & 3.142 & 0.000 & 3.142 & 1.571 & 0.663 & 10.996 \\
2.5 & 0.000 & 3.142 & 0.000 & 3.142 & 1.571 & 1.030 & 10.996 \\
5.0 & 0.000 & 3.142 & 0.000 & 3.142 & 1.571 & 1.560 & 10.996 \\
\hline
\end{tabular}
}
\label{tab:param-pre}
\end{table*}

\begin{table*}[hbtp]
\centering
\caption{ Post-optimized parameter specification. Parameters of the gates that appear in the post-optimization circuits of $\langle ZZ\rangle$, $\langle XZ\rangle$, and $\langle XX\rangle$ in Fig.~\ref{fig:pipeline}(c)(2)(i.). All parameters are in radians.}
\vspace{0.6em}
\footnotesize{
\begin{tabular}{ c | c | c | c | c | c | c | c | c }
$R\l\si{\angstrom}\r$ & $\phi^{zz}$ & $\phi'^{zz}$ & $\phi^{xz}_0$ & $\phi^{xz}_1$ & $\phi'^{xz}_1$ & $\phi^{xx}_0$ & $\phi^{xx}_1$ & $\phi'^{xx}_1$ \\
\hline
0.7 & 1.175 & 1.100 & 1.137 & 1.175 & 1.100 & 1.137 & 4.166 & 2.670 \\
0.85 & 1.571 & 1.464 & 1.571 & 1.571 & 1.464 & 1.571 & 4.499 & 3.035 \\
1.0 & 4.235 & 4.090 & 3.236 & 4.235 & 4.090 & 3.236 & 0.804 & 5.661 \\
1.1 & 1.477 & 1.301 & 2.469 & 1.477 & 1.301 & 4.712 & 4.266 & 2.871 \\
1.3 & 1.470 & 1.219 & 4.279 & 1.470 & 1.219 & 2.708 & 4.109 & 2.790 \\
1.6 & 0.088 & 5.975 & 4.141 & 0.088 & 5.975 & 2.570 & 2.438 & 1.263 \\
1.8 & 1.571 & 1.049 & 1.571 & 1.571 & 1.049 & 1.571 & 3.669 & 2.620 \\
2.0 & 1.571 & 0.911 & 1.571 & 1.571 & 0.911 & 1.571 & 3.393 & 2.482 \\
2.5 & 1.571 & 0.540 & 1.571 & 1.571 & 0.540 & 1.571 & 2.652 & 2.111 \\
5.0 & 1.571 & 0.013 & 1.571 & 1.571 & 0.013 & 1.571 & 1.596 & 1.583 \\
\hline
\end{tabular}
}
\label{tab:param-post}
\end{table*}

\begin{table*}[hbtp]
\centering
\caption{Post-optimized parameter specification. Parameters of the gates that appear in the post-optimization circuits of $\langle YY\rangle$ in Fig.~\ref{fig:pipeline}(c)(2)(ii.). All parameters are in radians.}
\vspace{0.6em}
\footnotesize{
\begin{tabular}{ c | c }
$R\l\si{\angstrom}\r$ & $\theta$ \\
\hline
0.7 & 0.075 \\    
0.85 & 0.106 \\
1.0 & 0.145 \\
1.1 & 0.175 \\
1.3 & 0.249 \\
1.6 & 0.396 \\
1.8 & 0.520 \\
2.0 & 0.663 \\
2.5 & 1.030 \\
5.0 & 1.560 \\
\hline
\end{tabular}
}
\label{tab:param-post2}
\end{table*}

\section*{Acknowledgements}
This work was supported as part of a joint development agreement between Dow and 1QBit. We are grateful to Peter Margl from Dow for technical discussions. The authors thank Marko Bucyk at 1QBit for his careful review and editing of the manuscript.

\bigskip
\section*{Data availability}
\revision{Atomic coordinates for the experimental points can be found in Supplementary Data 1.} All \revision{other} data that support the findings of this study are available from the corresponding author upon reasonable request.

\section*{Code availability}
The code is available from the corresponding author upon reasonable request.

\section*{Competing interests}
The authors declare no competing interest.

\section*{Author information}
\subsection*{Author Contributions}
Y.K. and T.Y. designed the entire research project. N.A. and A.Z. coordinated the research project. Y.K. and T.Y. designed the simulation aspects of the research. Y.K., E.L. and M.P.C performed the classical simulations. Y.K., E.L., M.P.C., L.H., and V.S. prepared the quantum circuits. Y.K., E.L., M.P.C., S.M., A.J.G., A.Z. and T.Y. analyzed the data of the simulations and hardware-related data. Y.N., S.J., and J.K. designed the hardware aspects of the research. Y.N. compiled and optimized the circuits for their execution on the hardware. A.O.M. and J.H.V.N. executed the circuits on the hardware. Y.N., A.O.M., and J.H.V.N. analyzed the hardware-specific data. Y.K., E.L., M.P.C., Y.N., S.M., A.J.G., S.J., N.A., A.Z., and T. Y. contributed to the writing of the manuscript.
\subsection*{Corresponding authors}
Correspondence to any of the following: Yukio Kawashima, Erika Lloyd, Marc P. Coons, Yunseong Nam, or Takeshi Yamazaki.


\def\bibsection{\section*{\refname}}
\bibliography{my_bib.bib}

\newpage
\begin{figure*}[hbtp]
\centering
\includegraphics[width=\textwidth]{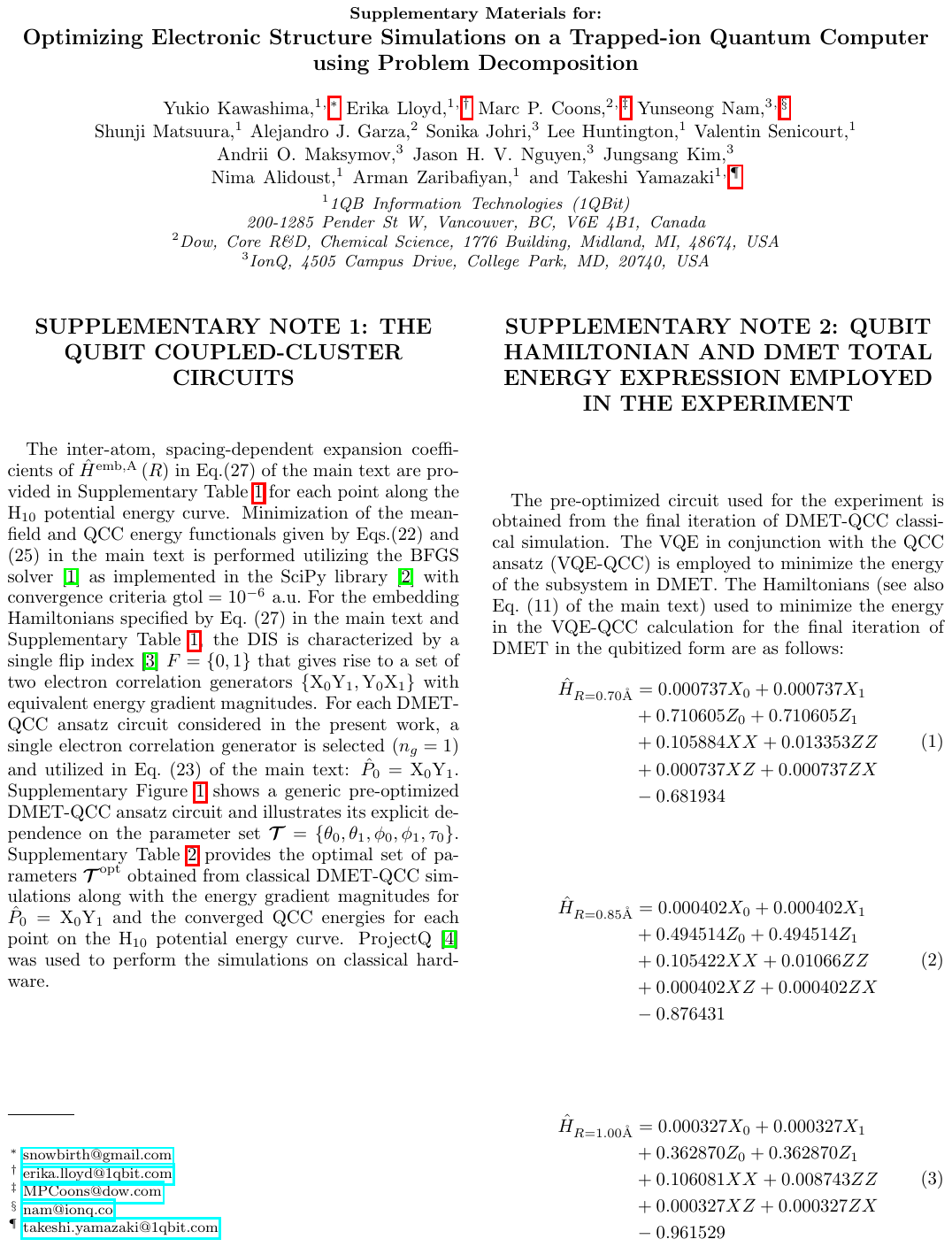}
\end{figure*}

\newpage
\begin{figure*}[hbtp]
\centering
\includegraphics[width=\textwidth]{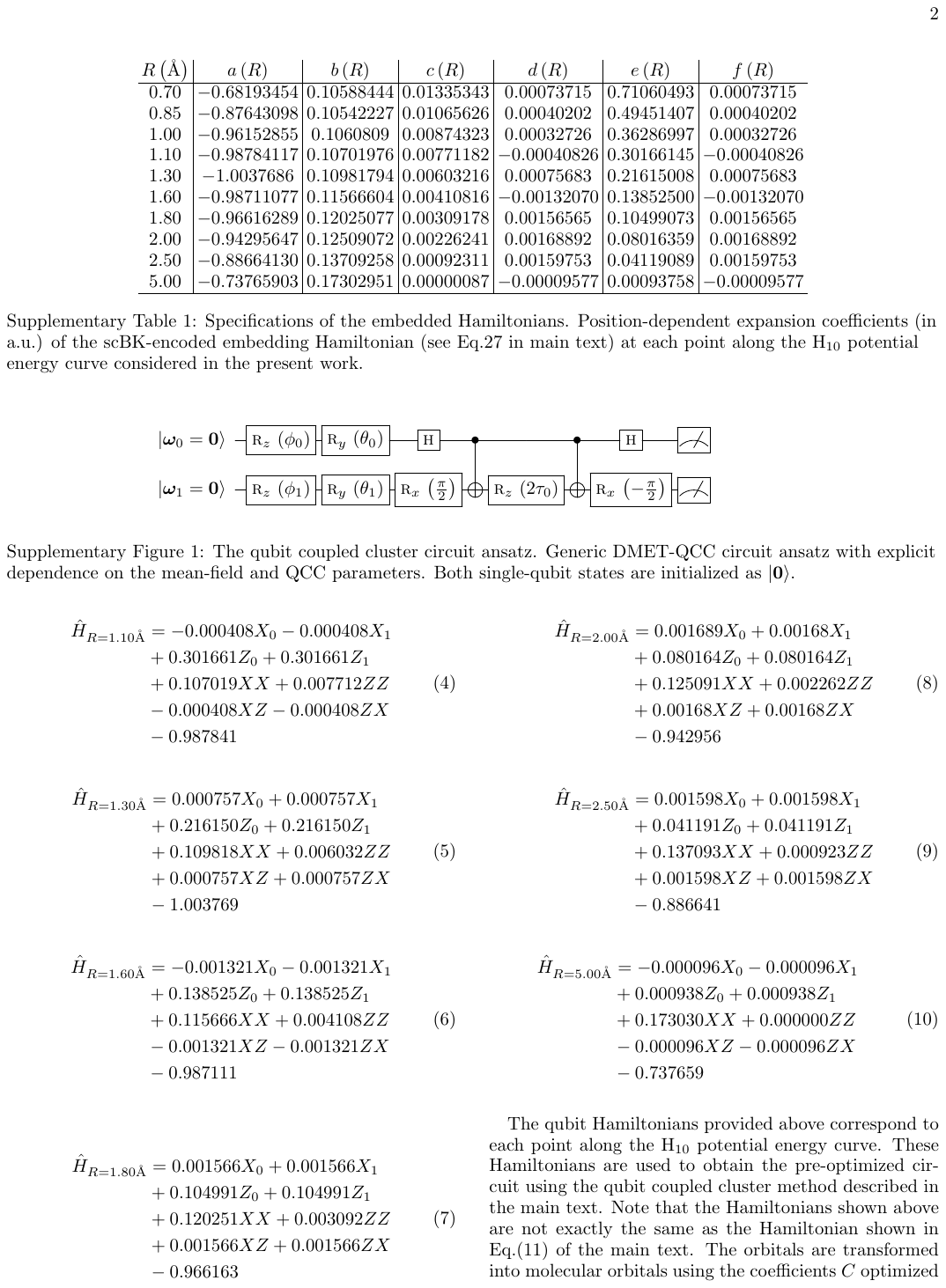}
\end{figure*}

\newpage
\begin{figure*}[hbtp]
\centering
\includegraphics[width=\textwidth]{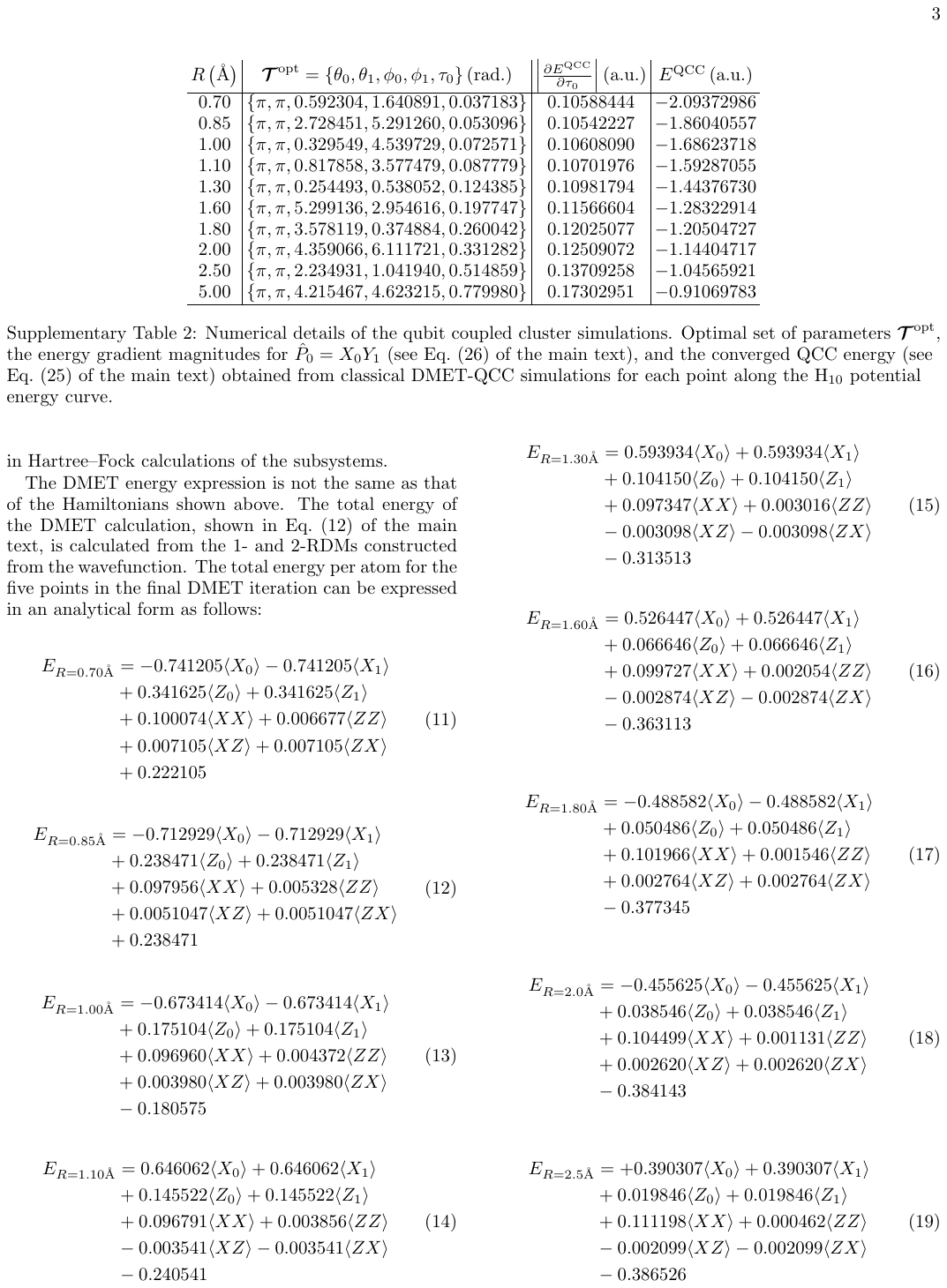}
\end{figure*}

\newpage
\begin{figure*}[hbtp]
\centering
\includegraphics[width=\textwidth]{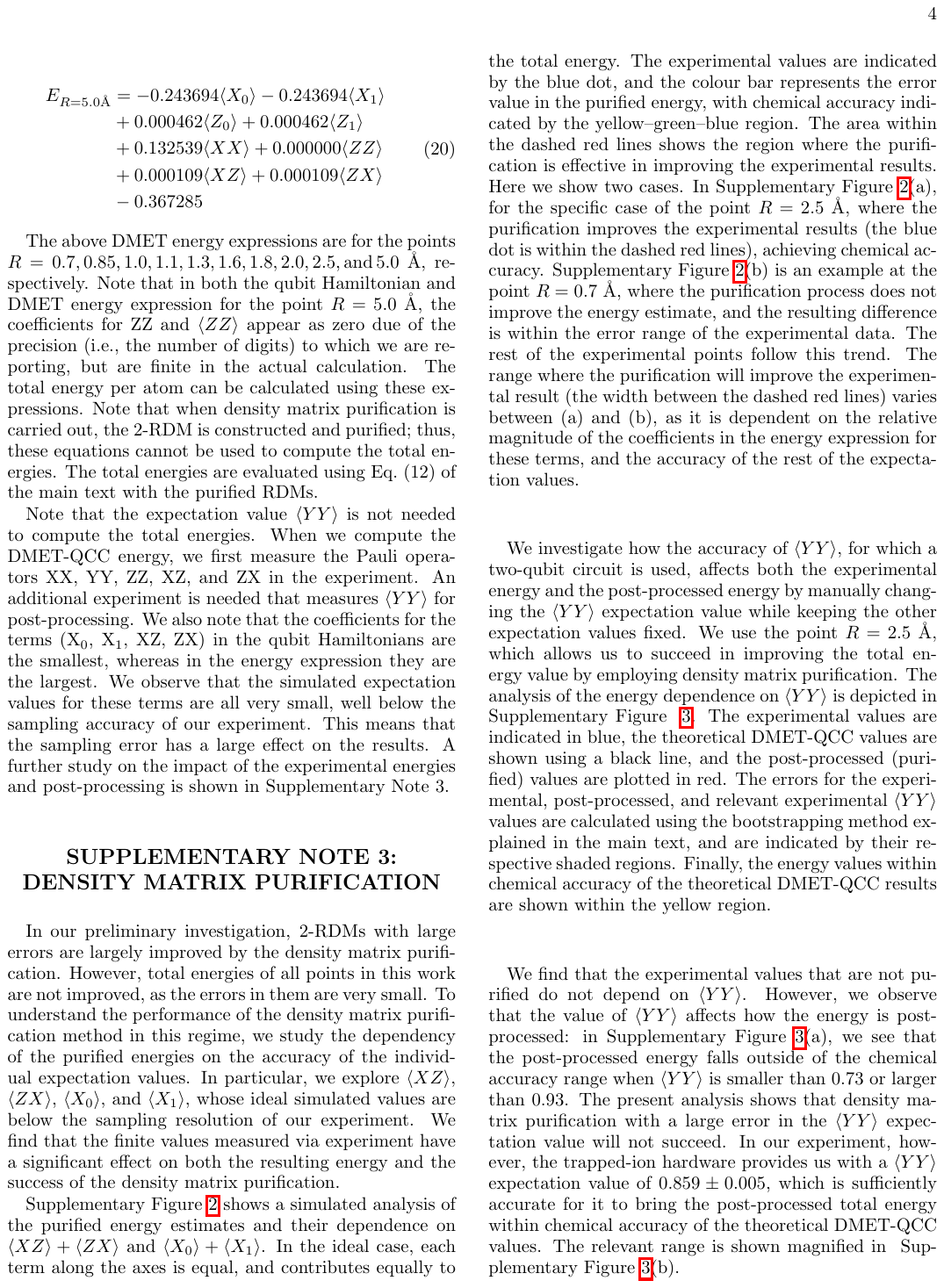}
\end{figure*}

\newpage
\begin{figure*}[hbtp]
\centering
\includegraphics[width=\textwidth]{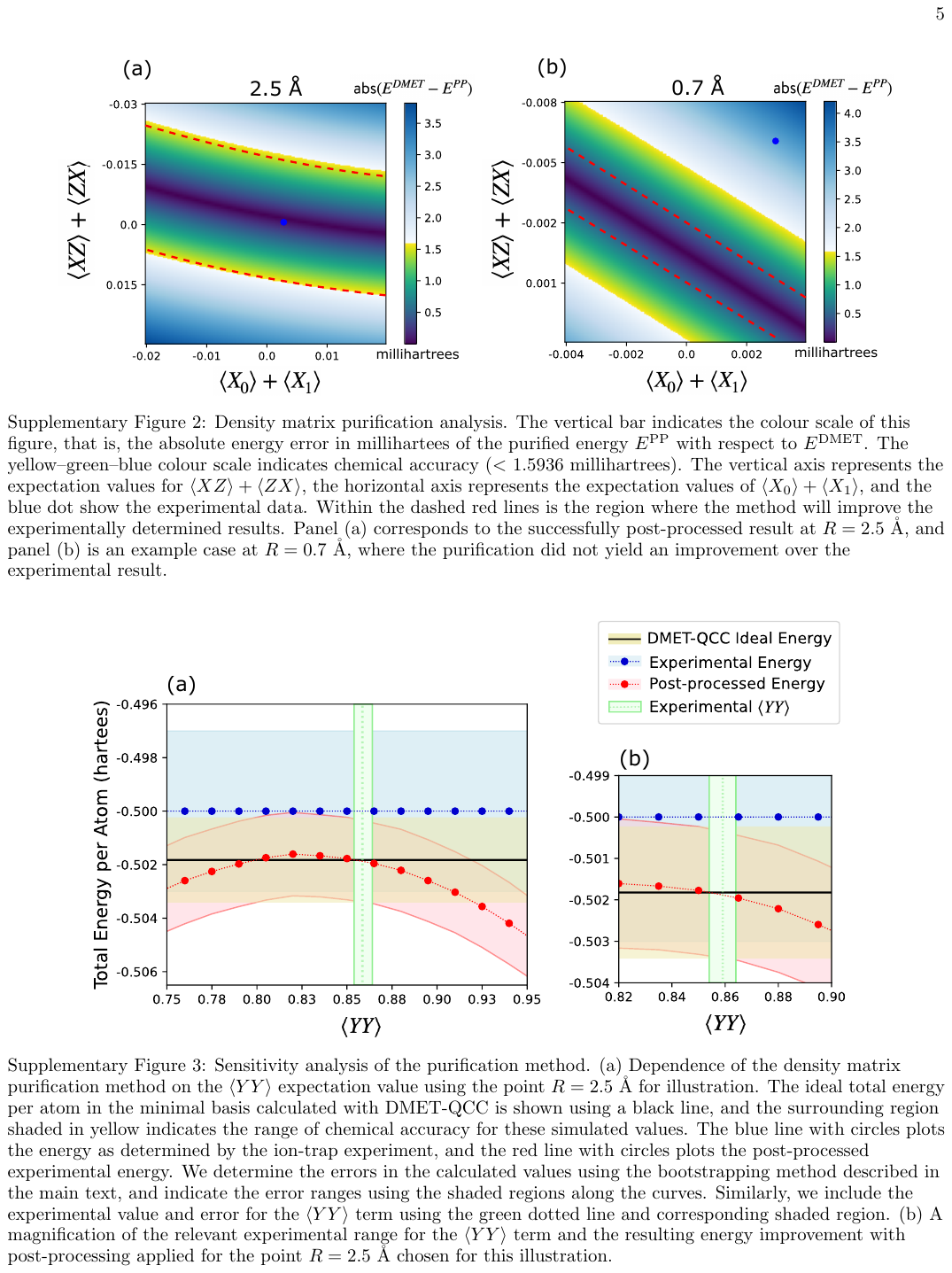}
\end{figure*}

\newpage
\begin{figure*}[hbtp]
\centering
\includegraphics[width=\textwidth]{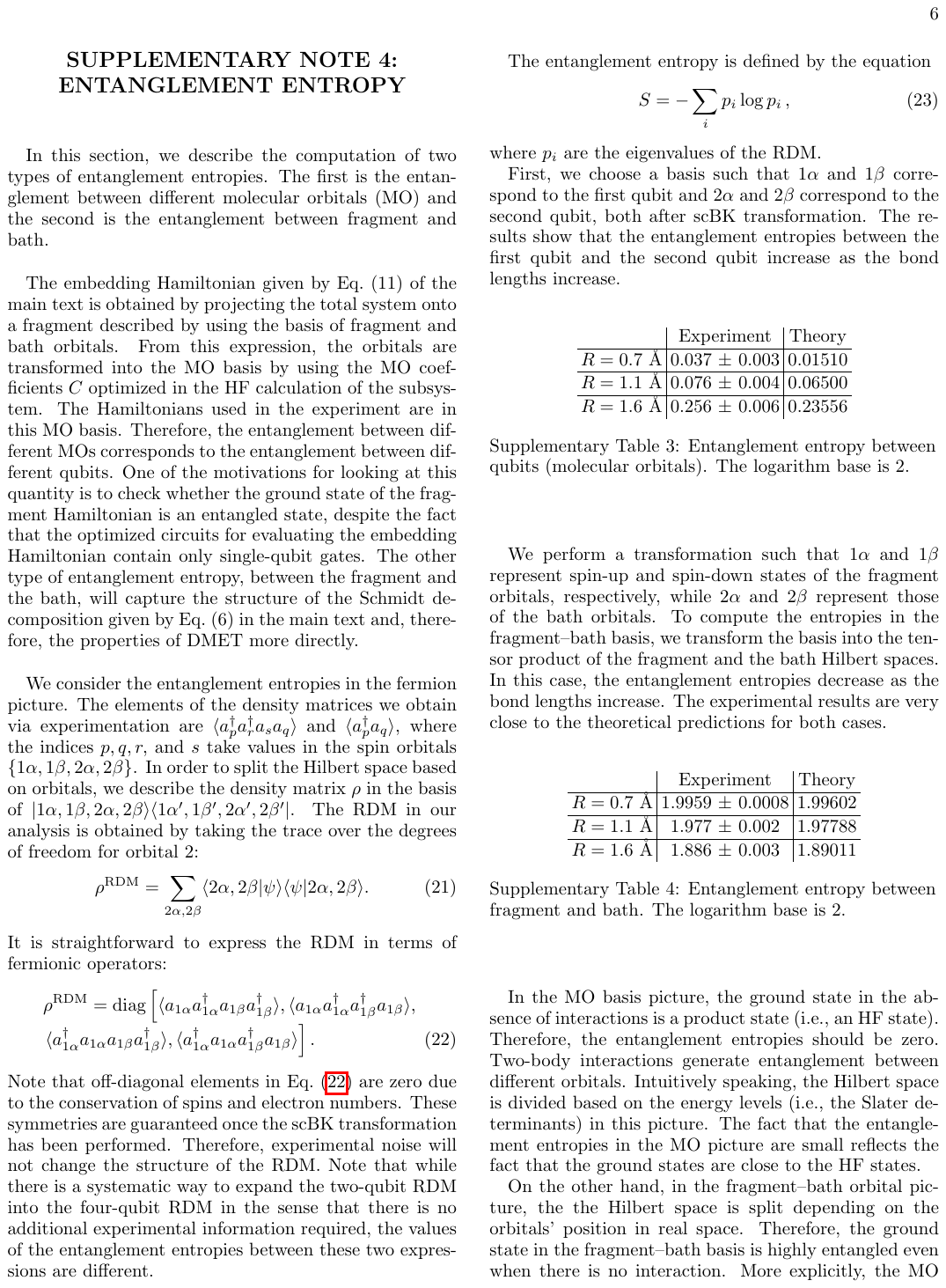}
\end{figure*}

\newpage
\begin{figure*}[hbtp]
\centering
\includegraphics[width=\textwidth]{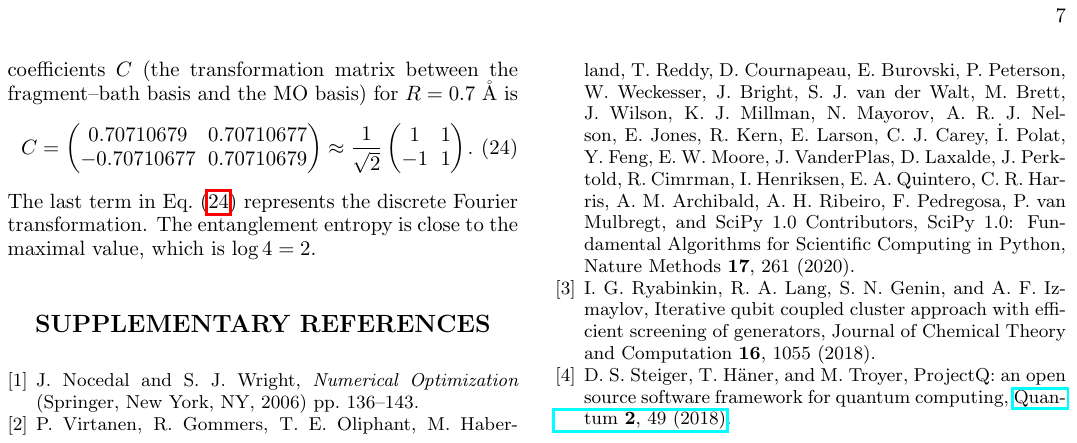}
\end{figure*}

\newpage
\begin{figure*}[hbtp]
\centering
\includegraphics[width=\textwidth]{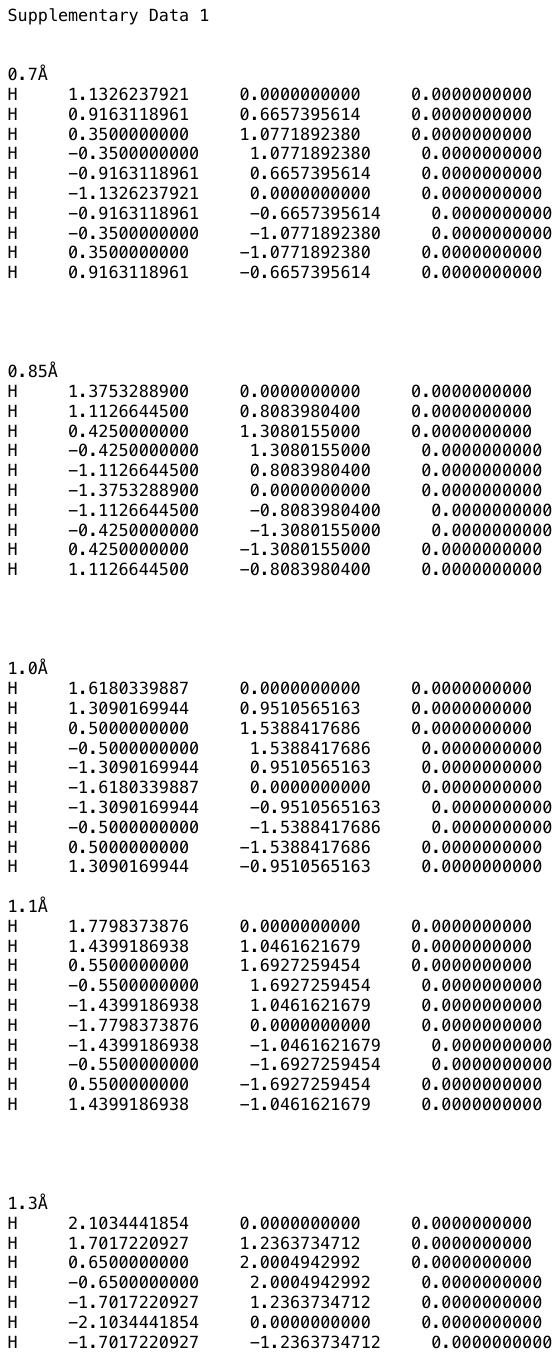}
\end{figure*}

\newpage
\begin{figure*}[hbtp]
\centering
\includegraphics[width=\textwidth]{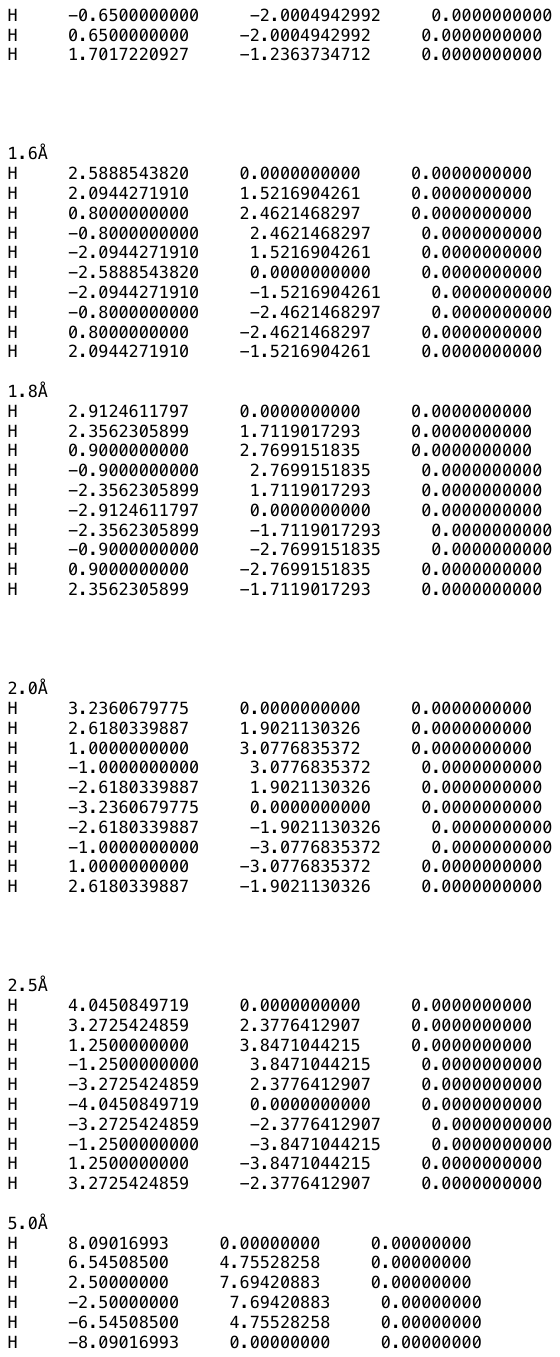}
\end{figure*}

\newpage
\begin{figure*}[hbtp]
\centering
\includegraphics[width=\textwidth]{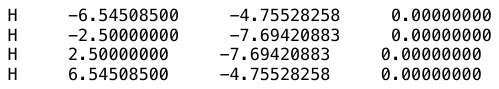}
\end{figure*}

\end{document}